\documentclass[twocolumn,pre]{revtex4}

\usepackage{dcolumn}
\usepackage{amsmath}
\usepackage{amssymb}
\usepackage{graphicx}
\usepackage{bm}
\usepackage[T1]{fontenc}
\usepackage{color}
\usepackage{url}
\usepackage[bf]{subfigure}

\begin{document}

\title{Seeking the best Internet Model}

\author{F.\ A.\ Rodrigues\, P.\ R.\ Villas Boas, G.\ Travieso and L. da F. Costa}
\affiliation{Instituto de F\'{\i}sica de S\~ao Carlos. Universidade de S\~ ao Paulo, S\~{a}o Carlos, SP, PO
Box 369, 13560-970, phone
+55 16 3373 9858,FAX +55 16 3371 3616, Brazil,\\
luciano@if.sc.usp.br}

\begin{abstract}

The models of the Internet reported in the literature are mainly aimed
at reproducing the scale-free structure, the high clustering
coefficient and the small world effects found in the real Internet,
while other important properties (e.g. related to centrality and
hierarchical measurements) are not considered. For a better
characterization and modeling of such network, a larger number of
topological properties must be considered.  In this work, we present a
sound multivariate statistical approach, including feature spaces and
multivariate statistical analysis (especially canonical projections),
in order to characterize several Internet models while considering a
larger set of relevant measurements. We apply such a methodology to
determine, among nine complex networks models, which are those most
compatible with the real Internet data (on the autonomous systems
level) considering a set of 21 network measurements. We conclude that
none of the considered models can reproduce the Internet topology with
high accuracy.

\end{abstract}


\maketitle

\section{Introduction}

In the Internet, an autonomous system (AS) is a large domain of IP
addresses that usually belongs to one organization such as a
university, a private company, or an Internet Service Provider. Since
AS are connected through border routers, the Internet can be
considered as consisting of interconnected AS. The understanding of
the fundamental mechanisms that govern the Internet evolution and
emergence are fundamental for modeling and simulating of dynamical
process, such as attacks~\cite{Albert2000eaa} and cascade
failures~\cite{Motter2004cca}, as well as for trying to improve
protocols and routing.

Large data sets about the Internet connections have been available
since the 90s. In 1999, Faloutsos et al.~\cite{Faloutsos1999plr}
showed that the distribution of connections is follow a power law,
despite the fact that new vertices and edges appear and disappear all
the time. This finding boosted the modeling and characterization of
the Internet. Among the obtained results, it has been shown that the
scale-free structure is important for providing network tolerance to
random failures~\cite{Albert2000eaa} and traffic
congestion~\cite{moreno2003cla,sole2001ita}. However, such a topology
makes the network vulnerable to intentional
attacks~\cite{cohen2001biu}. At the same time, the Internet protocol
efficiency is highly influenced by the network connectivity, while the
power law degree distribution results in an absence of an epidemic
threshold, which favors the spreading of computer
viruses~\cite{Pastorsatorras2001ess}.

The models proposed to generate the Internet topology vary from
completely random to those including preferential
attachments~\cite{Jin2000iit}. Accurate models for the Internet are
particularly important for growth forecast, architecture planning and
design, and to provide topologies for dynamical process
simulation. Although the characterization of the Internet structure is
becoming more and more precise, just a few models can statically
reproduce, and even so in approximate fashion, the Internet
evolution~\cite{yook2002mis}. While the current models are mainly
aimed at the degree distribution, other important features --- such as
those quantified by central and hierarchical measurements --- have not
teen considered in these models. This approach can result in
inaccurate and incomplete models. For instance, Alderson et
al.~\cite{Alderson2005tts} showed that networks with the same number
of vertices and edges, but distinct structure, can present the same
degree distribution (see also~\cite{Costa07}). In this way, the fact
that a model reproduces the same degree distribution as the real
network is not enough to validation. This suggests that most current
Internet models can be biased, undermining endeavors such as the
prediction of Internet evolution and dynamical simulations. In this
paper, we apply an alternative approach to determine the accuracy of
network models, by considering multivariate statistical analysis and
Bayesian decision theory~\cite{Duda2000pc, costa2001saa,
CostaAndrade07, Clauset06}.

Multivariate statistical methods have not been considered by complex
networks researchers until recently. The application of such methods
in classification of network has been suggested recently
(e.g.~\cite{Costa:survey,Clauset06,NewmanLeicht07}). Multivariate
statistical methods allow the consideration of a large set of
variables and can be of great help for network modeling. Indeed, a
model can be considered as being accurate if it can generate networks
whose structural properties --- quantified by a large set of network
measurements --- are statistically similar to those found for the real
network being considered.

In this work we present the application of multivariate statistical
methods, namely canonical projections and Bayesian decision theory, in
order to determine which among a set of Internet models is the most
appropriated to generate AS topologies. We considered nine different
complex networks models and a set of 21 measurement in our analysis.

\section{Concepts and Methods}

The considered Internet database, defined at the level of autonomous
systems (AS), is available at the web site of the National Laboratory
of Applied Network Research~(\texttt{http://www.nlanr.net}). The data
was collected in February $1998$, with the network containing $3522$
vertices and $6324$ edges. For the network characterization, we took
into account a set of 21 network measurements: (i) $\langle k
\rangle$, average vertex degree; (ii) $k_{max}$, maximum degree, (iii)
$\langle cc \rangle$, average clustering coefficient; (iv) $k_{nn}$,
average neighbor connectivity; (v) $\ell$, average shortest path
length; (vi) $r$, assortative coefficient; (vii) $\langle B \rangle$,
average betweenness, (viii) $c_D$, central point dominance; (ix)
$\mathit{st}$, straightness coefficient of the degree distribution;
(x) $\langle k_2 \rangle$, hierarchical degree of level two; (xi)
$\langle cc_{2}\rangle$, hierarchical clustering coefficient of level
two; (xii) $cv_2$, convergence ratio of level two; (xiii) $dv_2$,
divergence ratio of level two; (xiv) $E_2$, average inter-ring degree
of level two; (xv) $A_2$, average intra-ring degree of level two; (xi)
$\langle k_3
\rangle$, hierarchical degree of level three; (xvii) $\langle cc_{3}\rangle$, hierarchical clustering
coefficient of level three; (xviii) $cv_3$, convergence ratio of level
three; (xix) $dv_3$, divergence ratio of level three; (xx) $E_3$,
average inter-ring degree of level three; and (xxi) $A_3$, average
intra-ring degree of level three. The classification was obtained by
considered canonical variable analysis and Bayesian decision
theory~\cite{Duda2000pc,costa2001saa,Costa:survey}.

\subsection{Network measurements}

The AS network can be represented in terms of its adjacency matrix $A$, whose elements $a_{ij}$ are equal to
one whenever there is a connection between the vertices $i$ and $j$, or equal to 0, otherwise. The average
vertex degree is given as
\begin{equation}
  \langle k \rangle = \frac{1}{N}\sum_{ij} a_{ij}.
\end{equation}
The clustering coefficient of a node $i$ ($cc_i$) is defined by the proportion of links between the vertices
within its neighborhood, $l_i$, divided by the number of links that could possibly exist between them
($k_i(k_i-1)/2$). The average clustering coefficient is computed as
\begin{equation}
  \langle cc \rangle = \frac{1}{N}\sum_{i=1}^N cc_i = \frac{1}{N}\sum_{i=1}^N \frac{\sum_{j=1}^N \sum_{m=1}^N
a_{ij}a_{jm}a_{mi} }{k_i(k_i -1)}.
  \label{clustercoeff2}
\end{equation}
The average neighbor connectivity ($k_{nn}$) measures the average degree of vertices neighbor of the each
vertex in the network~\cite{Pastorsatorras2001dac}. The average shortest path length ($\ell$) is calculated
by taking into account the shortest distance between each pair of vertices in the network. The assortative
coefficient measures the correlation between vertex degrees,i.e.,
\begin{equation}
  r = \frac{
            \frac{1}{M} \sum_{j>i}k_i k_j a_{ij} -
            \left[ \frac{1}{M} \sum_{j>i}
                   \frac{1}{2} (k_i+k_j) a_{ij} \right]^2
           }
           {
            \frac{1}{M}\sum_{j>i}\frac{1}{2}(k_i^2+k_j^2) a_{ij} -
            \left[ \frac{1}{M}\sum_{j>i}
                   \frac{1}{2}(k_i+k_j) a_{ij}  \right]^2
           }.
  \label{pearson}
\end{equation}
The straightness coefficient ($st$) quantifies the level to which a
log-log distribution of points approaches a power law, which is
computed in terms of the Pearson correlation coefficient of the loglog
degree distribution~\cite{Costa:survey}.

The considered centrality measurements are based on the betweenness
centrality, which is defined as
\begin{equation}
  B_u = \sum_{ij} \frac{\sigma(i,u,j)}{\sigma(i,j)},
  \label{betweenness}
\end{equation}
where $\sigma(i,u,j)$ is the number of shortest paths between vertices
$i$ and $j$ that pass through vertex $u$, $\sigma(i,j)$ is the total
number of shortest paths between $i$ and $j$, and the sum is over all
pairs $i,j$ of distinct vertices. The average betweenness centrality
($\langle B\rangle$) is computed considering the whole set of vertices
in the network. The central point dominance is defined in terms of the
betweenness by the following equation,
\begin{equation}
  \mathit{c_D} = \frac{1}{N-1}\sum_{i}(B_{\mathrm{max}} - B_i).
  \label{eq:cpd}
\end{equation}
where $B_{\mathrm{max}}$ represents the maximum betweenness found in
the network.

Complex networks measurements can also be defined in a hierarchical (or concentric) way~\cite{Costa2004hbc,
Costa:2005a,Costa2006gac,CostaAndrade07}, i.e. by considering the successive neighborhoods around each node.
Therefore, it is interesting to define the ring of vertices $R_d(i)$, which is formed by those vertices
distant $d$ edges from the reference vertex $i$. The hierarchical degree at distance $d$ ($k_d(i)$) is
defined as the number of edges connecting the rings $R_d(i)$ and $R_{d+1}(i)$. The hierarchical clustering
coefficient is given by the number of edges in the respective $d$-ring ($m_{d}(i)$), divided by the total
number of possible edges between the vertices in that ring, i.e.,
\begin{equation}
  cc_{d}(i) = \frac{2 m_{d}(i)}{n_{d}(i)(n_{d}(i)-1)},
\end{equation}
where $n_{d}(i)$ represents the number of vertices in the ring
$R_d(i)$. The convergence ratio at distance $d$ of $i$ corresponds to
the ratio between the hierarchical degree at distance $d-1$ and the
number of vertices in the ring $R_d(i)$,
\begin{equation}
  \mathit{cv}_d(i) = \frac{k_{d-1}(i)}{n_d(i)}.
  \label{cratio}
\end{equation}
The divergence ratio corresponds to the reciprocal of the convergence
ratio, i.e.,
\begin{equation}
  \mathit{dv}_d(i) = \frac{n_d(i)}{k_{d-1}(i)}.
  \label{dratio}
\end{equation}
Finally, the average inter ring degree is given by the average of the
number of connections between each vertex in the ring $R_d(i)$ and
those in $R_{d+1}(i)$,
\begin{equation}
E_d(i) = \frac{k_d(i)}{n_d(i)};
\end{equation}
and the average intra ring degree is defined as the average among the degrees of the vertices in the ring
$R_d(i)$,
\begin{equation}
A_d(i) = \frac{2m_d(i)}{n_d(i)},
\end{equation}
The average of each hierarchical measurements is obtained by taking
into account the local hierarchical measurement of each vertex in the
network.

\subsection{Network models}

The following nine complex network types are considered for modeling
the Internet:
\begin{enumerate}

  \item \emph{Erd\H{o}s-R\'{e}nyi random graph (ER}): The network is
  constructed connecting each pair of vertices in the network with a
  fixed probability $p$~\cite{Bollobas98}, where each pair of vertices
  $(i,j)$ is selected at random only once. This model generates a
  Poisson degree distribution.

  \item \emph{Small-world model of Watts and Strogatz (WS)}: To
  construct this small-word network, one starts with a regular lattice
  of $N$ vertices in which each vertex is connected to $\kappa$
  nearest neighbors in each direction.  Each edge is then randomly
  rewired with probability $p$~\cite{Watts98:Nature}.

  \item \emph{Waxman geographical Internet model~(WGM)}: Geographical
  networks can be constructed by distributing $N$ vertices at random
  in a 2D space and connecting them according to the distance. The
  model suggested by Waxman to model the Internet
  topology~\cite{Waxman1988rmc} considers the probability to connect
  two vertices $i$ and $j$, distant $D_{ij}$, as $P(i\rightarrow j)
  \sim \theta e^{-\lambda D_{ij}}$.

  \item \emph{Barab\'{a}si-Albert scale-free model~(BA)}: The network
  is generated by starting with a set of $m_0$ vertices and, at each
  time step, the network grows with the addition of a new vertice with
  $m$ links. The vertices which receive the new edges are chosen
  following a linear preferential attachment rule, i.e.\ the
  probability of the new vertex $i$ to connect with an existing vertex
  $j$ is proportional to the degree of $j$, $\mathcal{P}(i\rightarrow
  j) = k_j/\sum_u k_u$~\cite{Barabasi99:Science}.

  \item \emph{Limited scale-free model~(LSF)}: The network is
  generated as in the BA model but the maximum degree is limited in
  order to be equal to the degree of the real network~\cite{amaral2000csw}.

  \item \emph{Scale-free model of Dorogovtsev, Mendes and
  Samukhin~(DMS)}: This network is constructed as in the BA model, but
  the preferential attachment rule is defined as
  $\mathcal{P}(i\rightarrow j) = (k_j+k_0)/\sum_u (k_u+k_0)$
  ~\cite{dorogovtsev2000sgn}. The constant $k_0$ controls the initial
  attractiveness and provides variation of connectivity from
  $-m<k_0<\infty$, allowing a larger variation in the exponent of the
  power law, $\gamma = 3 + k_0/m$ (for the BA model, $\gamma = 3)$.

  \item \emph{Nonlinear scale-free network model~(NLSF)}: The network
  is constructed as in the BA model, but instead of a linear
  preferential attachment rule, the vertices are connected following a
  nonlinear preferential attachment rule, i.e., $P_{i\rightarrow j} =
  k_j^\alpha/\sum_u k_u^\alpha$. In this case, while for $\alpha < 1$,
  the network has a stretched exponential degree distribution, for
  $\alpha > 1$ a single site connects to nearly all other
  sites~\cite{krapivsky2001ogr}.

  \item \emph{The geographic directed preferential Internet topology
  model~(GdTang)}: This internet generator constructs direct AS
  networks by considering some rules of the BA model. At each time
  step, a new vertex $i$ and $m$ edges are added to the network. The
  new vertex $i$ connects with a vertex $j$ according to the the rule
  $P_{i \rightarrow j} = k_j^{out}/\sum_u k_u^{out}$. The remaining
  $m-1$ edges connect any vertex in the network according to the rule:
  the outgoing endpoint of each edge (node $i$) is chosen with
  probability $P_i = k_j^{in}/\sum_u k_u^{in}$ and the incoming
  endpoint (node $j$) with $P_j = k_j^{out}/\sum_u k_u^{out}$. With
  probability $\beta$, the added edge is local and the endpoints are
  restricted to the same region. The nodes are spatially distributed
  considering a pre-defined distribution. On the other hand, with
  probability $1-\beta$, the edge is global and can connect any
  endpoints. With probability $p$, each added edge may become a
  undirected edge~\cite{Bar2005gdp}.

  \item \emph{The Inet internet topology generator}: The Inet 3.0 has
  been based on the AS growth analysis since November 1997.
  Basically, this model assumes an exponential growth rate of the
  number of AS and it is computed the number of months $t$ necessary
  to obtain a network with $N$ vertices. Next, the out-degree
  frequency and the rank out-degree distribution are calculated. A
  fraction of $n$ vertices are assigned to degree one and the
  remaining vertices are assigned out-degrees according to the
  out-degrees frequency. More details about this model can be found
  in~\cite{Jin2000iit,Winick2002iit}.

\end{enumerate}

The models (iv)-(ix) produce networks with power law degree
distributions as observed in the Internet. The models (i)-(iii) are
considered in the current network classification because of their
ability to reproduce network topological properties such as the small
world effect and the high average clustering coefficient values. The
NLSF model is simulated considering the exponents of the preferential
attachment equal to $\alpha = 0.5$ and $\alpha = 1.5$. The models WGM,
GdTang and Inet were developed specifically to generate Internet
topologies. Despite GdTang generates directed networks, we symmetrize
the connections --- directed connections were transformed in
undirected. This transformation does not alters the network structure.
All considered networks were formed by $N = 3522$ vertex and the
average vertex degree adjusted to that of the original network
($\langle k_{AS} \rangle = 3.59$).

\subsection{Classification methodology}

A multivariate statistical method was adopted in order to associate
(through classification) the Internet to the most likely among the
considered models~\cite{Costa:survey}. The classification was obtained
by associating the real network to the model which best reproduces its
topology, as quantified by the measurements. The features space was
defined for 10 classes (the nonlinear model is defined considering two
different exponents for the preferential attachment). For each model,
50 networks were generated and 21 measurements were computed. In this
way, each network model realization was represented by a feature
vector composed by 21 elements in the space of attributes. Such a
space was projected into 2D by using canonical variable
analysis~\cite{Costa:survey,Johnson:book} and the region of
classification was obtained by Bayesian decision
theory~\cite{Duda2000pc,costa2001saa}.

Canonical analysis has been used to reduce the dimensionality of the
measurement feature space. It provides a powerful extension of
principal component analysis~\cite{Johnson:book}, performing
projections which optimize the separation between known categories of
objects. To perform the canonical analysis it is necessary to
construct a matrix which quantifies the variation inside the groups
previously defined, and a second matrix which quantifies the variation
among these groups. If we consider $C$ classes (network models), each
one identified as $C_i, i = 1, \dots, C$, and that each network
realization $n$ is represented by its respective feature vector
$\vec{x}_n = (x_1,x_2,\dots,x_p)^T$, the intraclass scatter
matrix is defined as
\begin{equation}\label{canonical1}
S_{\mathrm{intra}} = \sum_{i=1}^C \sum_{n \in C_i} \left ( \vec{x_n} - \vec{\left< x \right>}_i \right )
\left ( \vec{x_n} - \vec{\left< x \right>}_i \right )^T,
\end{equation}
and the interclass scatter matrix is given as,
\begin{equation}\label{canonical2} S_{\mathrm{inter}} =
\sum_{i=1}^{C} N_i \left ( \vec{\left< x \right>}_i - \vec{\left< x \right>} \right ) \left ( \vec{\left< x
\right>}_i - \vec{\left< x \right>} \right )^T,
\end{equation}
where $\vec{\left< x \right>}_i$ corresponds to the average of a given
variable for the class $i$ and $\vec{\left< x \right>}$ is the general
average of a given variable for all classes.

By computing the eigenvectors of the matrix
$S_{\mathrm{intra}}^{-1}S_{\mathrm{inter}}$ and selecting those
corresponding to highest absolute eigenvalues, $\lambda_1,\dots,
\lambda_p$, it is possible to project the set of variables into less dimension ---usually $2$ or $3$ dimensions, depending on
the number of highest eigenvalues considered~\cite{costa2001saa}.

The Bayesian decision is performed in order to obtain the regions of
classification by considering non-parametric
estimation~\cite{Duda2000pc}. In this, case the mass probabilities
$P_i$, which corresponds to the probability that an network belongs to
class $C_i$, as well as the conditional probability densities,
$p(\vec{x_n} | C_i)$, are estimated by using non-parametric methods
(see~\cite{costa2001saa,Duda2000pc}). The Bayes rule can then be
expressed as:
\begin{eqnarray*}
\mathrm{if} \: f(\vec{x}_n|C_a)P(C_m) = \mathrm{max}_{b=1,m}\{f(\vec{x}_n|C_b)P(C_b)\} \;\\ \mathrm{then} \:
\mathrm{select} \: C_a ,
\end{eqnarray*}
where $\vec{x}_n$ is the vector that stores the network set of
measurements and $C_a$ is the class of networks associated to the
model $a$. Further details about such an approach are discussed
in~\cite{Costa:survey}.

\section{Results and discussion}

The network models were generated while considering parameters that
best approximate the average vertex degree and/or the average
clustering coefficient of the real network. In this way, we considered
fpr each model: (i) ER, $p = \langle k_{AS} \rangle/(N-1)$; (ii) SW,
$\kappa
\simeq \langle k_{AS} \rangle/2 = 2$ and $p = 1 - [\langle cc_{AS}
\rangle(4\kappa-2)/(3\kappa - 3)]^{1/3}$; (iii) BA, $m \simeq \langle
k_{AS} \rangle/2 = 2$; (iv) WGM, the parameters $\lambda = 1.35$ and
$\theta = 1$ were adjusted in order to obtain a degree similar to the
real network; (iv) LSF, $m \simeq \langle k_{AS} \rangle/2 = 2$ and
the maximum degree was taken equal to that observed in the real
network; (v) DMS, $m \simeq \langle k_{AS} \rangle/2 = 2$ and $k_0 =
m(\gamma_{AS} - 3)$, where $\gamma_{AS} = 2.2$ is the exponent of the
degree distribution of Internet~\cite{Pastorsatorras2001dac}; (vi) KP,
$m \simeq \langle k_{AS} \rangle/2 = 2$ and the coefficient of the
nonlinearity was taken $\alpha = 0.5$ and $\alpha = 1.5$; (vii)
GdTang, $p = 0.5$ and $\beta = 0.07$; and (viii) Inet 2.0, the
fraction of vertices with degree equal to one was defined as observed
in the Internet. The measurements $\langle k_{AS} \rangle$ and
$\langle cc_{AS} \rangle$ are the average degree and the average
clustering coefficient found in the Internet, respectively. For each
model, 50 networks were generated and a set of 21 different
measurements were computed for each one (nine non-hierarchical and 6
hierarchical, where the hierarchical measurements consider the second
and third hierarchies).

Table~\ref{Tab:Measurements} presents the five most commonly used
measurements for network characterization.  According to their values,
we may conclude that the Inet 3.0 is the most accurate model, in spite
of $\langle cc \rangle = 0$. However, such a set of measurements does
not quantify the majority of network properties and a larger set of
measurements must be considered in order to enhance the precision
of the analysis.

In order to obtain the classification of the Internet by using
canonical variable analysis and Bayesian decision theory, according to
the set of models and measurements, we took into account the following
eight measurements configurations:
\begin{enumerate}
  \item $\{k_{max}, \ell, r \}$.
  \item $\{ \langle k\rangle, k_{max},  \langle cc\rangle, \ell, r, c_D\}$
  \item $\{\langle cc\rangle,k_{nn},\ell,c_D,st\}$
  \item $\{k_{max},\langle cc\rangle, k_{nn}, \ell, r, \langle B \rangle, st \}$
  \item $\{k_{nn},\ell,r,\langle B\rangle,\langle k_2\rangle,\langle cc_2\rangle, \langle k_3\rangle, \langle cc_3 \rangle \}$
  \item $\{\langle k\rangle, k_{max},\langle cc\rangle,\ell,r,c_D,\langle k_2\rangle,\langle cc_2\rangle \}$
  \item $\{\langle k_2\rangle,\langle cc_2\rangle, \langle cv_2\rangle, \langle E_2\rangle,\langle A_2\rangle,
  \langle k_3 \rangle,\langle cc_3, \rangle \langle cv_3 \rangle$, $\langle E_3 \rangle, \langle A_3\rangle \}$
  \item $\{ \langle k \rangle, k_{max}, \langle cc \rangle, k_{nn}, \ell, r, \langle B \rangle, c_D,
  st, \langle k_2 \rangle, \langle cc_{2}\rangle, cv_2$, $E_2, A_2, \langle k_3 \rangle, \langle cc_{3}\rangle, cv_3, E_3, A_3\}$.
\end{enumerate}

\begin{table*}[t]
\begin{center}
\caption{Average and standard deviation of the average degree ($\langle k \rangle$), maximum degree
($k_{max}$), average clustering coefficient ($\langle cc \rangle$),
average shortest path length ($\ell$), assortative coefficient (r) and
the central point dominance ($c_D$). Each measurements was computed
taking into account 50 realizations of each model.}
\begin{tabular}{l|c|c|c|c|c|c}
\hline Network  &$\langle k \rangle$ & $k_{max}$ &$\langle cc \rangle$ &$\ell$ &$r$ &$c_D$\\\hline

Internet            &3.59          &742        &0.19               &3.46          &-0.21          &0.33 \\
ER                  &3,59$\pm0,04$ &12$\pm$2 &0.0010$\pm$0.0005  &6.5$\pm$0.1   &-0.004$\pm$0.0140 &0.010$\pm$0.02\\
SW                  &4$\pm$0       &7$\pm$21   &0.37$\pm$0.05      &10$\pm$1      &-0.023$\pm$0.01 &0.020$\pm$ 0.05\\
BA                  &4$\pm$0       &158$\pm$33 &0.010$\pm$0.002    &4.4$\pm$0.1   &-0.06$\pm$0.01 &0.20$\pm$0.05\\
WGM                 &3.35$\pm$0.05 &11$\pm$2 &0.12$\pm$0.01      &18$\pm$1      &0.15$\pm$0.1     &0.15$\pm$0.05\\
LSF                 &4$\pm$0       &160$\pm$33 &0.010$\pm$0.002    &4.3$\pm$0.1   &-0.06$\pm$0.01 &0.20$\pm$0.05\\
DMS                 &4$\pm$0       &352$\pm$63 &0.04$\pm$0.01      &3.7$\pm$0.1   &-0.13$\pm$0.01 &0.30$\pm$0.05\\
LNSF($\alpha=0.5$)  &4$\pm$0       &40$\pm$5   &0.002$\pm$0.001    &5.1$\pm$0.1   &0.08$\pm$0.01 &0.08$\pm$0.02\\
NLSF($\alpha=1.5$)  &4$\pm$0       &1500$\pm$300 &0.3$\pm$0.1      &2.7$\pm$0.2   &-0.25$\pm$0.05 &0.6$\pm$0.1\\
GdTang              &4.7$\pm$0.1   &380$\pm$50   &0.03$\pm$0.01    &5.0$\pm$0.1   &-0.18$\pm$0.01 &0.10$\pm$0.01\\
Inet 3.0            &3,23$\pm$0    &771$\pm$0    &0$\pm$0          &3,38$\pm$0.01 &-018$\pm$0.01 &0.52$\pm$0.02\\
\hline
\end{tabular}
\label{Tab:Measurements}
\end{center}
\end{table*}

\begin{figure*}[t]
\begin{center}
\subfigure[]{\includegraphics[width=0.48\linewidth]{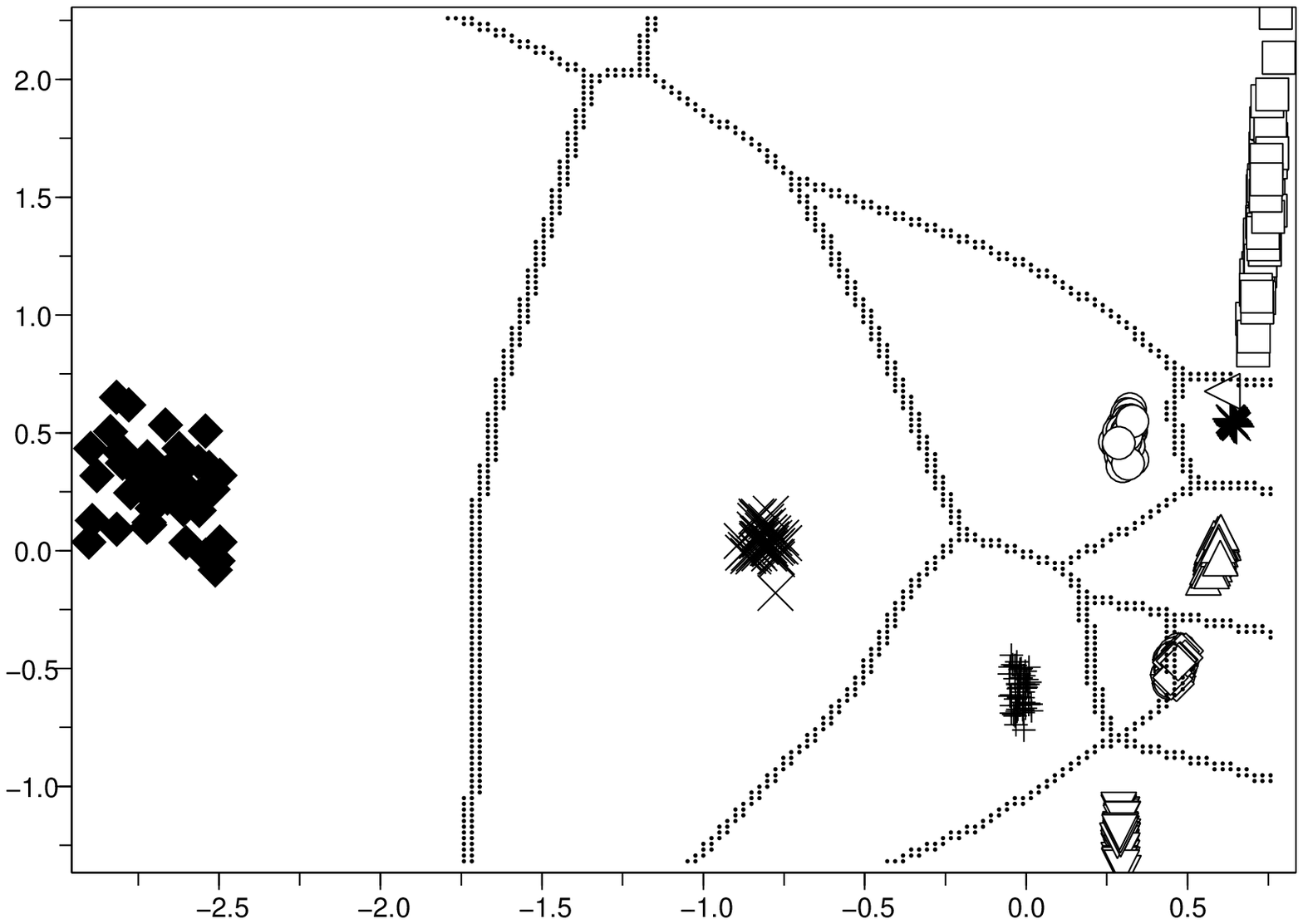}}
\subfigure[]{\includegraphics[width=0.48\linewidth]{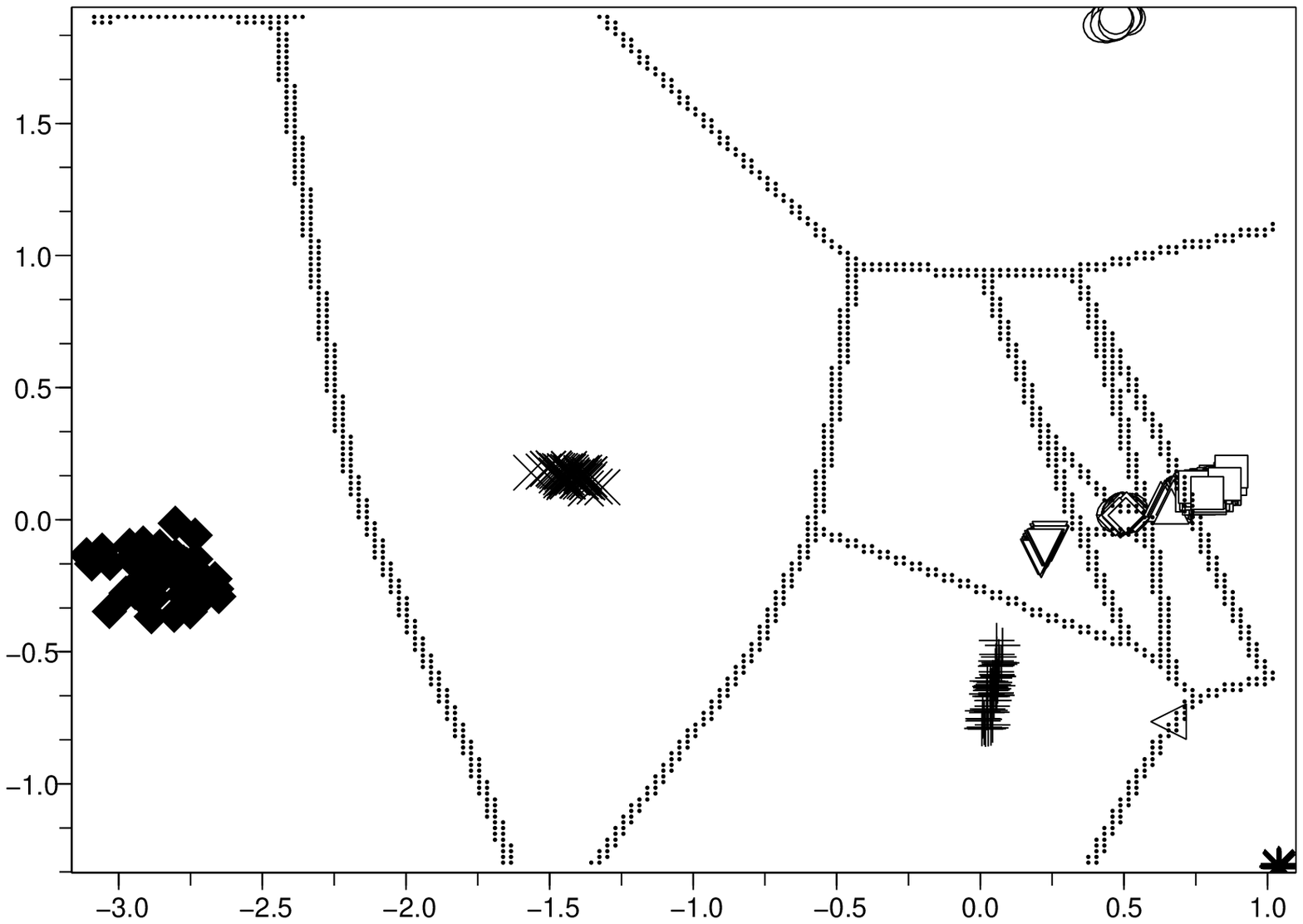}}
\subfigure[]{\includegraphics[width=0.48\linewidth]{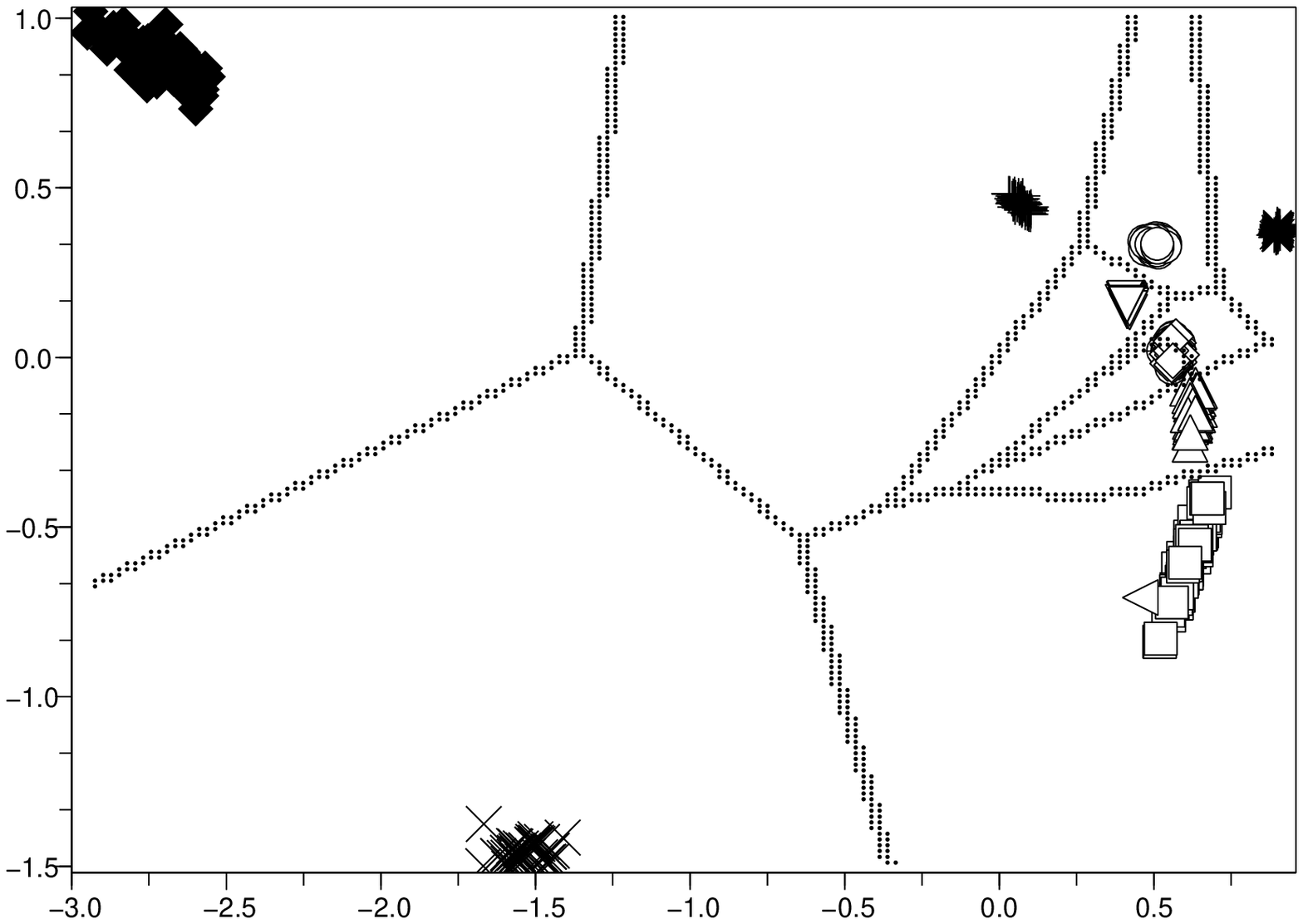}}
\subfigure[]{\includegraphics[width=0.48\linewidth]{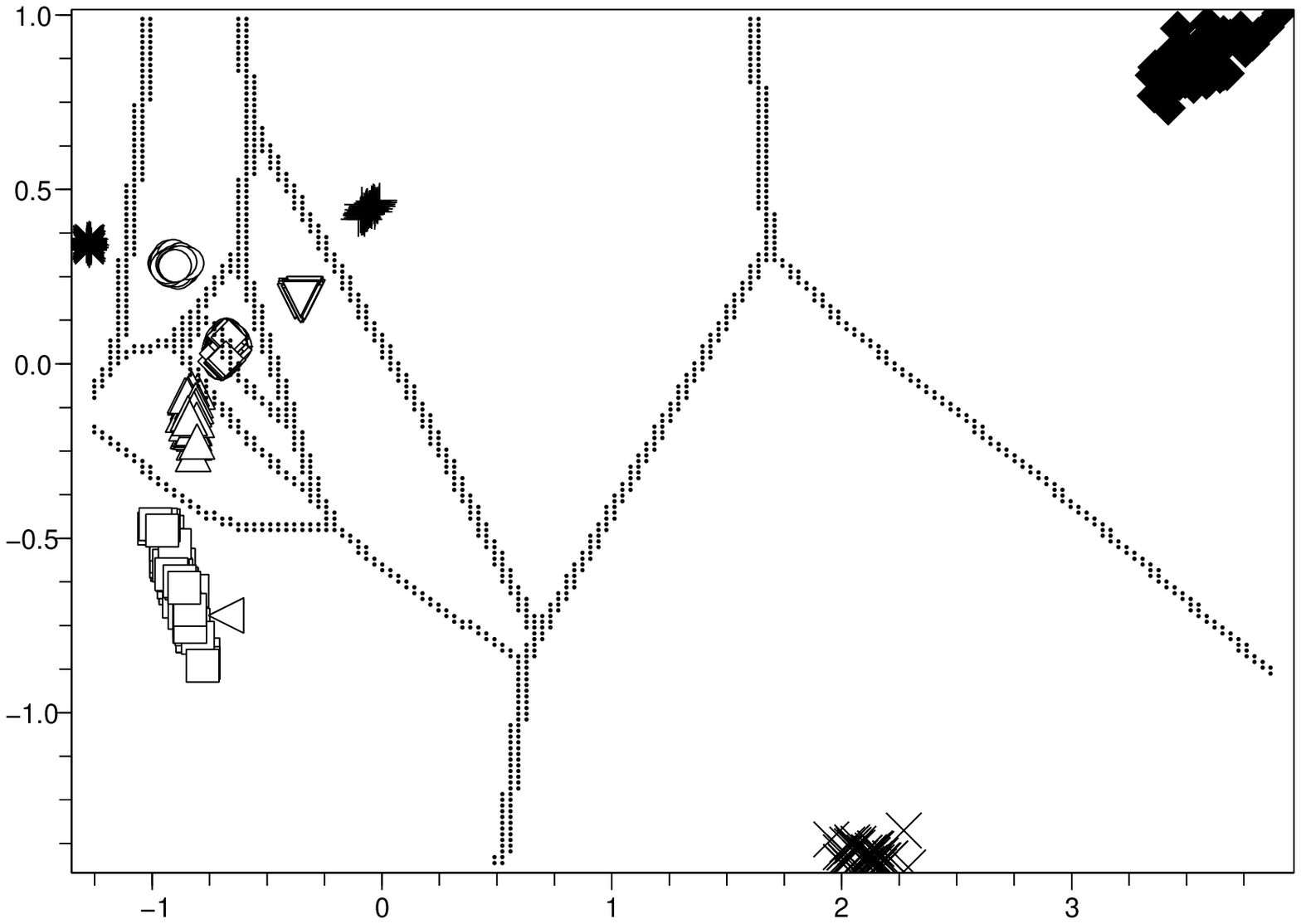}}
\end{center}
 \caption{Classification obtained considering different set of
 measurements. The network realizations are represented by dots,
 corresponding to the following models: $+$ ER, $\times$ WS, $\oplus$
 BA, $\blacklozenge$ WGM, $\lozenge$ LSF, $\vartriangle$ DMS,
 $\triangledown$ NLSF ($\alpha = 0.5$), $\square$ NLSF ($\alpha =
 1.5$), $\circ$ GdTang and $\ast$ Inet 3.0. The real network is
 represented by $\triangleleft$. The set of measurements in each case
 are (a) $\{k_{max}, \ell, r \}$, (b)~$\{ \langle k\rangle, k_{max},
 \langle cc\rangle, \ell, r, c_D\}$, (c) $\{\langle
 cc\rangle,k_{nn},\ell,c_D,st\}$ and (d) $\{k_{max},\langle cc\rangle,
 k_{nn}, \ell, r, \langle B \rangle, st \}$ }
 \label{Fig:classification1}
\end{figure*}

\begin{figure*}[t]
\begin{center}
\subfigure[]{\includegraphics[width=0.48\linewidth]{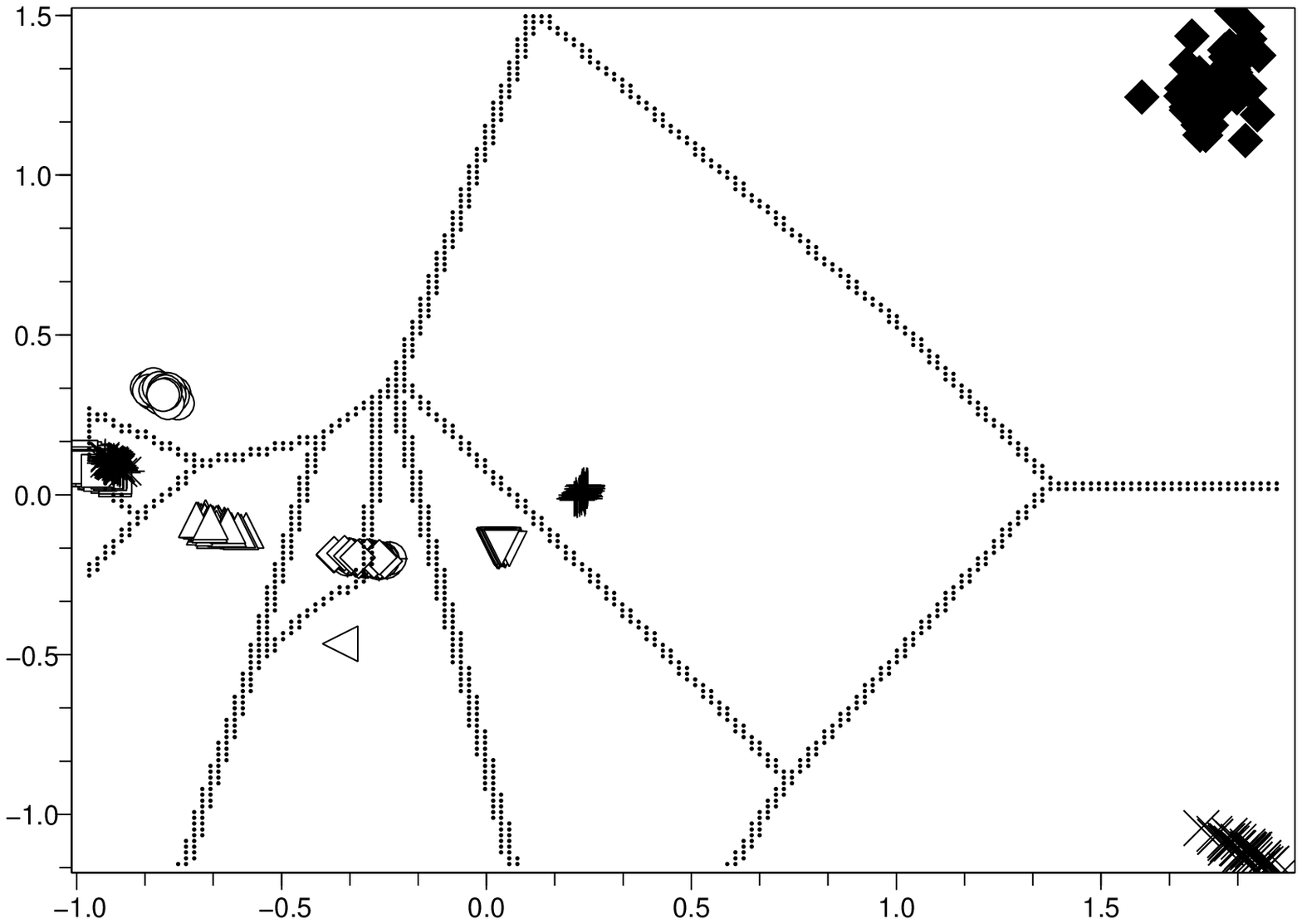}}
\subfigure[]{\includegraphics[width=0.48\linewidth]{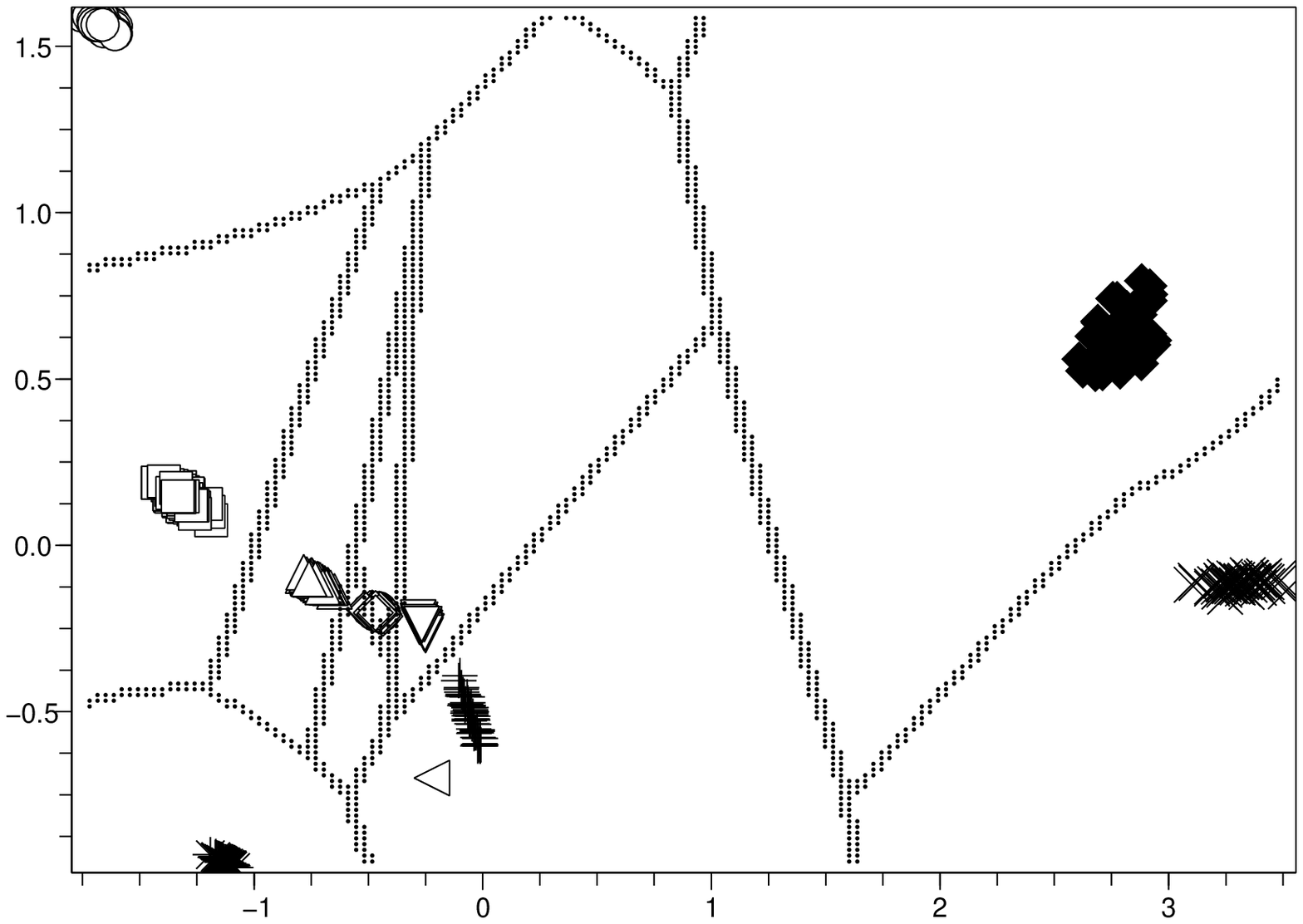}}
\subfigure[]{\includegraphics[width=0.48\linewidth]{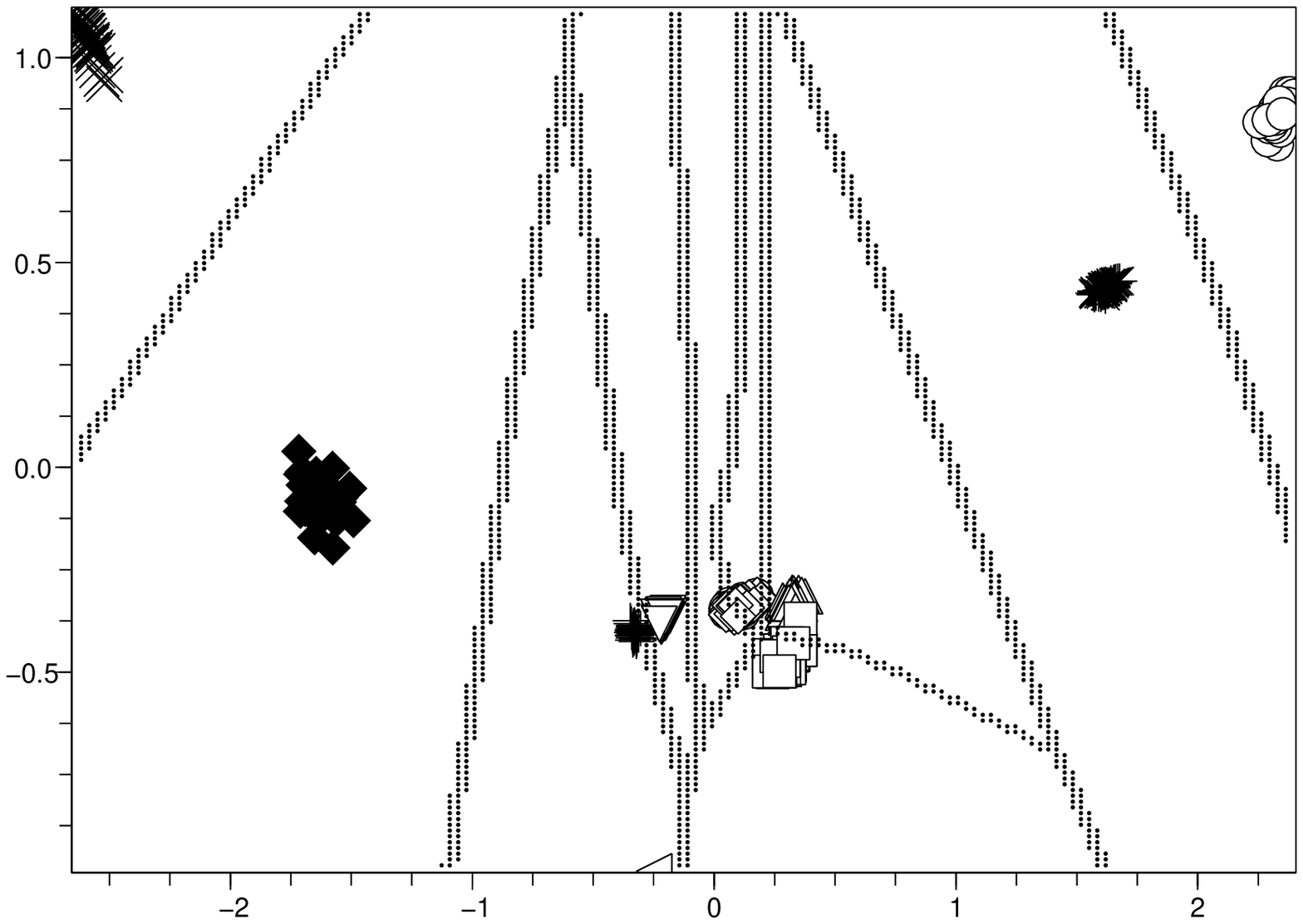}}
\subfigure[]{\includegraphics[width=0.48\linewidth]{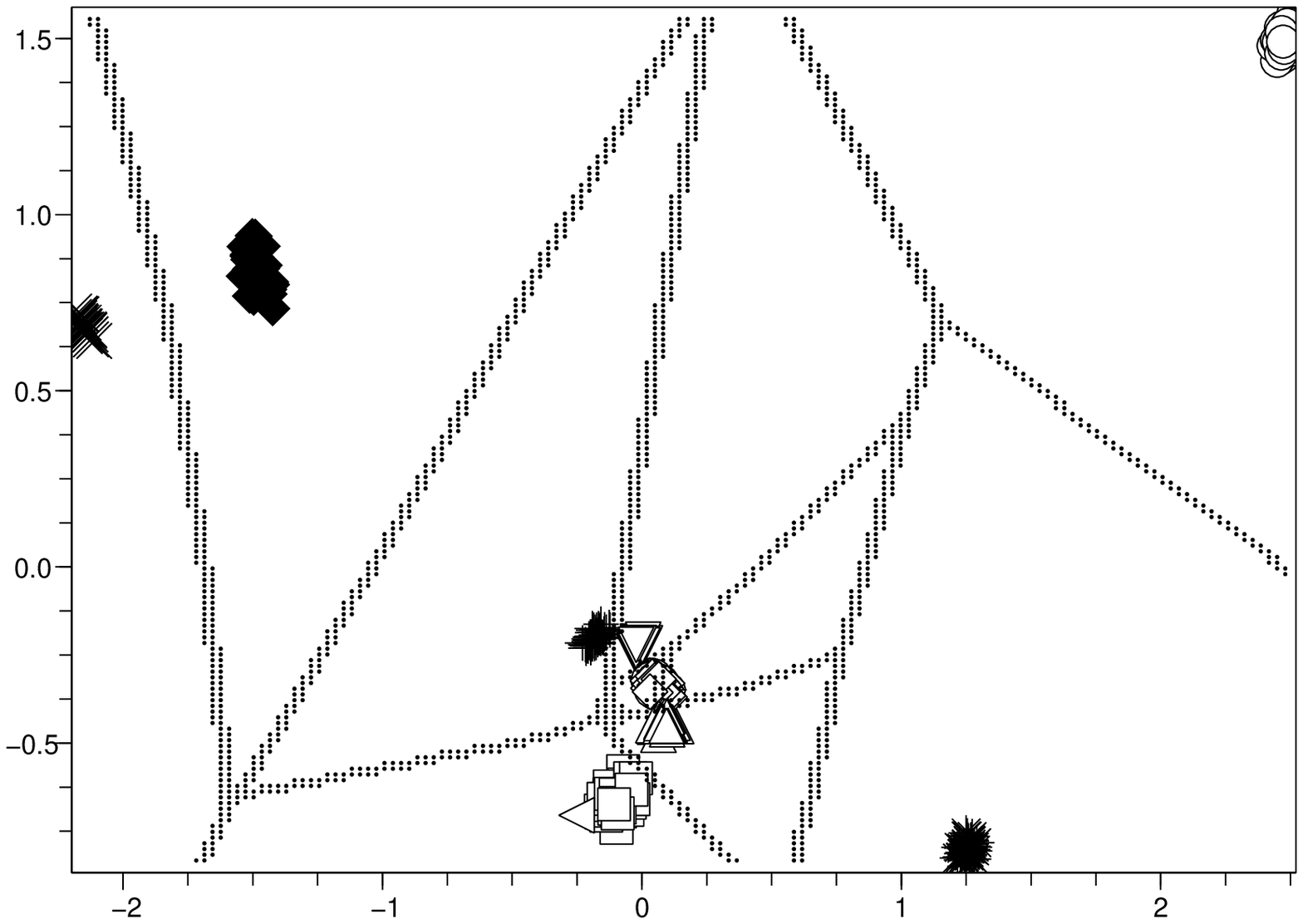}}
\end{center}

\caption{Classification obtained considering different sets of  measurements.
The network realizations are represented by dots, corresponding to the
following models: $+$ ER, $\times$ WS, $\oplus$ BA, $\blacklozenge$
WGM, $\lozenge$ LSF, $\vartriangle$ DMS, $\triangledown$ NLSF ($\alpha
= 0.5$), $\square$ NLSF ($\alpha = 1.5$), $\circ$ GdTang and $\ast$
Inet 3.0. The real network is represented by $\triangleleft$. The set
of measurements in each case are (a) $\{k_{nn},\ell,r,\langle
B\rangle,\langle k_2\rangle,\langle cc_2\rangle, \langle k_3\rangle,
\langle cc_3 \rangle \}$, (b) $\{\langle k\rangle, k_{max},\langle
cc\rangle,\ell,r,c_D,\langle k_2\rangle,\langle cc_2\rangle \}$, (c)
$\{\langle k_2\rangle,\langle cc_2\rangle, \langle cv_2\rangle,
\langle E_2\rangle,\langle A_2\rangle, \langle k_3 \rangle,\langle
cc_3, \rangle \langle cv_3
\rangle$, $\langle E_3 \rangle, \langle A_3\rangle \}$ and (d) $\{
\langle k \rangle, k_{max}, \langle cc
\rangle, k_{nn}, \ell, r, \langle B \rangle, c_D, st, \langle k_2
\rangle, \langle cc_{2}\rangle, cv_2$, $E_2, A_2,
\langle k_3 \rangle, \langle cc_{3}\rangle, cv_3, E_3, A_3\}$. }

  \label{Fig:classification2}
\end{figure*}

Figures~\ref{Fig:classification1} and~\ref{Fig:classification2}
present the obtained partitions and classifications. As we can see,
different classifications were obtained depending on the set of
measurements considered. For the set (i) and (ii)
(Figure~\ref{Fig:classification2}(a) and
~\ref{Fig:classification2}(b)), the Internet was best represented by
the model Inet 3.0. Indeed, this result is observed in
Table~\ref{Tab:Measurements} and reflects the biased classification
when a reduced set of measurements is considered. The Inet reproduces
well some topological measurements ($\langle k\rangle, k_{max}, \ell,
r$), while other measurements ( $\langle c\rangle$ and $c_D$) tend to
diverge. When the sets (vi) and (vii) are taken into account, the
Internet is best modeled by the ER network model
(Figures~\ref{Fig:classification2}(b)
and~\ref{Fig:classification2}(c)). This classification was not
expected, since ER model produces networks with topology different
from the Internet (see Table~\ref{Tab:Measurements}). In case the
measurements (iii), (iv) and (viii) are considered, the Internet was
classified as KP($\alpha=1.5$) (Figures~\ref{Fig:classification1}(c),
\ref{Fig:classification1}(d) and
\ref{Fig:classification2}(d)). Indeed, this model considers the non-linear preferential attachment, which has
been considered in other Internet models, such as that developed by
Zhou and Mondragon~\cite{zhou2004ami} --- which was not considered
here because it is suitable to reproduce only CAIDA
networks~\cite{Zhou:05}. For the set of measurements (v), the Internet
was classified as BA model, even if the BA model did not produce
assortative networks with high average clustering coefficient and
degree distribution with the same exponent as observed in Internet
($\gamma_{Ba} = 3$ and $\gamma_{AS} = 2.2$). In none of the
classifications, the real network was placed among the points that
defined each class. All these results suggest that none of the models
can reproduce the Internet topology with high accuracy. The ER, BA,
NLSF ($\alpha=1.5$) and Inet 3.0 can reproduce just some topological
properties of the real network. Therefore, such models can be
considered as roughly approximated. For a given model to reproduce the
Internet structure with precision, whatever the set of measurements
considered, the network would have to be classified as corresponding
to this model. Our results suggest that a revision of Internet
modeling must be considered in order to obtaining improved
prototypes. A possibility to obtain a better model of Internet is to
observe which of the properties of the ER, BA, NLSF and Inet 3.0 are
important for Internet evolution. In this case, a hybrid model may be
constructed.

\section{Conclusions}

In this work we presented an application of multivariate statistical
analysis to determine, among a set of pre-defined complex networks
models, which of them is potentially most suitable to represent the
Internet topology. Our results suggest that none of the considered
models reproduce all considered features of the Internet. Even models
developed specifically to reproduce the Internet structure --- such as
the Inet, WGM and GdTang --- do not seem to be very accurate. In order
to obtain more precise modeling, hybrid models can be constructed,
considering properties of the ER, BA, NLSF and Inet 3.0 that are
important for Internet evolution, as these models were the only that
reproduced, partially, some Internet topological properties.

The present work suggests that a revision in Internet modeling, which
can be assisted by the methods considered in this work. Also, it is
possible to extend our approach by considering the contribution of
each measurement for the separation in the phase space as a systematic
methodology for identifying the incompleteness of the models. This
approach can result in incremental improvements, allowing to quantify
the importance of each measurement in discrimination. The extension of
the modeling methods for other types of complex networks, such as
social and biological, is straightforward.

\section{Acknowledgments}

Luciano da F. Costa is grateful to FAPESP (proc. 05/00587-5), CNPq
(proc. 301303/06-1) for financial support.  Francisco A. Rodrigues
acknowledges FAPESP sponsorship (proc. 07/50633-9) and Paulino
R. Villas Boas is grateful to CNPq (141390/2004-2).

\section*{References}
\bibliographystyle{unsrt}
\bibliography{paper}

\begin{thebibliography}{10}

\bibitem{Albert2000eaa}
R.~Albert, H.~Jeong, and A.L. Barabasi.
\newblock Error and attack tolerance of complex networks.
\newblock {\em Nature}, 406(6794):378--382, 2000.

\bibitem{Motter2004cca}
A.E. Motter.
\newblock Cascade control and defense in complex networks.
\newblock {\em Physical Review Letters}, 93(9):98701, 2004.

\bibitem{Faloutsos1999plr}
M.~Faloutsos, P.~Faloutsos, and C.~Faloutsos.
\newblock {On power-law relationships of the Internet topology}.
\newblock {\em Proceedings of the conference on Applications, technologies,
  architectures, and protocols for computer communication}, pages 251--262,
  1999.

\bibitem{moreno2003cla}
Y.~Moreno, R.~Pastor-Satorras, A.~Vazquez, and A.~Vespignani.
\newblock {Critical load and congestion instabilities in scale-free networks}.
\newblock {\em Europhysics Letters}, 62(2):292--298, 2003.

\bibitem{sole2001ita}
R.V. Sol{\'e} and S.~Valverde.
\newblock {Information transfer and phase transitions in a model of internet
  traffic}.
\newblock {\em Physica A: Statistical Mechanics and its Applications},
  289(3-4):595--605, 2001.

\bibitem{cohen2001biu}
R.~Cohen, K.~Erez, D.~ben Avraham, and S.~Havlin.
\newblock {Breakdown of the Internet under Intentional Attack}.
\newblock {\em Physical Review Letters}, 86(16):3682--3685, 2001.

\bibitem{Pastorsatorras2001ess}
R.~Pastor-Satorras and A.~Vespignani.
\newblock Epidemic spreading in scale-free networks.
\newblock {\em Physical Review Letters}, 86(14):3200--3203, 2001.

\bibitem{Jin2000iit}
C.~Jin, Q.~Chen, and S.~Jamin.
\newblock {Inet: Internet topology generator}.
\newblock {\em University of Michigan Technical Report CSE-TR-433-00}, 2000.

\bibitem{yook2002mis}
S.H. Yook, H.~Jeong, and A.L. Barabasi.
\newblock {Modeling the Internet's large-scale topology}.
\newblock {\em Proceedings of the National Academy of Sciences},
  99(21):13382--13386, 2002.

\bibitem{Alderson2005tts}
D.~Alderson, J.C. Doyle, L.~Li, and W.~Willinger.
\newblock Towards a theory of scale-free graphs: Definition, properties, and
  implications.
\newblock {\em Internet Math}, 2(4):431--523, 2005.

\bibitem{Costa07}
L.~da~F.~Costa.
\newblock Seeking for simplicity in complex networks.
\newblock arXiv:physics/0702102v1, 2007.

\bibitem{Duda2000pc}
R.O. Duda, P.E. Hart, and D.G. Stork.
\newblock {\em {Pattern Classification}}.
\newblock Wiley-Interscience, 2000.

\bibitem{costa2001saa}
L.D.F. Costa.
\newblock {\em {Shape Analysis and Classification: Theory and Practice}}.
\newblock CRC Press, 2001.

\bibitem{CostaAndrade07}
L.~da~F.~Costa and R.~F.~S. Andrade.
\newblock What are the best hierarchical descriptors for complex networks?
\newblock arXiv:0705.4251v1, 2007.

\bibitem{Clauset06}
A.~Clauset, C.~Moore, and M.~E.~J. Newman.
\newblock Structural inference of hierarchies in networks,.
\newblock In {\em Proc. 23rd International Conference on Machine Learning
  (ICML)}, New York, 2006. Association of Computing Machinery.

\bibitem{Costa:survey}
L.~da~F. Costa, F.~A. Rodrigues, G.~Travieso, and P.~R.~Villas Boas.
\newblock Characterization of complex networks: A survey of measurements.
\newblock {\em Advances in Physics}, 56(1):167 -- 242, 2007.

\bibitem{NewmanLeicht07}
M.~E.~J. Newman and E.~A. Leicht.
\newblock Mixture models and exploratory data analysis in networks.
\newblock {\em Proc. Natl. Acad. Sci. USA}, 104:9564--9569, 2007.

\bibitem{Pastorsatorras2001dac}
R.~Pastor-Satorras, A.~V{\'a}zquez, and A.~Vespignani.
\newblock Dynamical and correlation properties of the internet.
\newblock {\em Physical Review Letters}, 87(25):258701, 2001.

\bibitem{Costa2004hbc}
L.F. Costa.
\newblock The hierarchical backbone of complex networks.
\newblock {\em Physical Review Letters}, 93(9):98702, 2004.

\bibitem{Costa:2005a}
L.~da~F.~Costa and F.~N. Silva.
\newblock Hierarchical characterization of complex networks.
\newblock {\em Journal of Statistical Physics}, 125(4):841--872, 2006.

\bibitem{Costa2006gac}
L.~da~F.~Costa and L.~E.~C. da~Rocha.
\newblock A generalized approach to complex networks.
\newblock {\em The European Physical Journal B-Condensed Matter},
  50(1):237--242, 2006.

\bibitem{Bollobas98}
B.~Bollob\'{a}s.
\newblock {\em Modern Graph Theory}.
\newblock Graduate Texts in Mathematics, Springer-Verlag, New York, 1998.

\bibitem{Watts98:Nature}
D.~J. Watts and S.~H. Strogatz.
\newblock Collective dynamics of small-world networks.
\newblock {\em Nature}, 393(6684):440--442, 1998.

\bibitem{Waxman1988rmc}
BM~Waxman.
\newblock {Routing of multipoint connections}.
\newblock {\em Selected Areas in Communications, IEEE Journal on},
  6(9):1617--1622, 1988.

\bibitem{Barabasi99:Science}
A.L. Barab\'asi and R.~Albert.
\newblock Emergence of scaling in random networks.
\newblock {\em Science}, 286:509, 1999.

\bibitem{amaral2000csw}
L.A.N. Amaral, A.~Scala, M.~Barthelemy, and HE~Stanley.
\newblock {Classes of Small-World Networks}.
\newblock {\em Proceedings of the National Academy of Sciences of the United
  States of America}, 97(21):11149--11152, 2000.

\bibitem{dorogovtsev2000sgn}
SN~Dorogovtsev, JFF Mendes, and AN~Samukhin.
\newblock {Structure of Growing Networks with Preferential Linking}.
\newblock {\em Physical Review Letters}, 85(21):4633--4636, 2000.

\bibitem{krapivsky2001ogr}
PL~Krapivsky and S.~Redner.
\newblock {Organization of growing random networks}.
\newblock {\em Physical Review E}, 63(6):66123, 2001.

\bibitem{Bar2005gdp}
S.~Bar, M.~Gonen, A.~Wool, and S.~Bar.
\newblock A geographic directed preferential internet topology model.
\newblock {\em Modeling, Analysis, and Simulation of Computer and
  Telecommunication Systems, 2005. 13th IEEE International Symposium on}, pages
  325--328, 2005.

\bibitem{Winick2002iit}
J.~Winick and S.~Jamin.
\newblock {Inet-3.0: Internet topology generator}.
\newblock {\em University of Michigan Technical Report CSE-TR-456-02}, 2002.

\bibitem{Johnson:book}
R.~A. Johnson and D.~W. Wichern.
\newblock {\em Applied Multivariate Statistical Analysis}.
\newblock Prentice Hall, fourth edition, 1998.

\bibitem{zhou2004ami}
S.~Zhou and R.J. Mondrag{\'o}n.
\newblock {Accurately modeling the internet topology}.
\newblock {\em Physical Review E}, 70(6):66108, 2004.

\bibitem{Zhou:05}
S.~Zhou, G.~Zhang, and G.~Zhang.
\newblock The chinese internet as-level topology.
\newblock arXiv:cs/0511101v4, 2005.

\end{thebibliography}

\end{document}